\documentclass[manuscript]{aastex63}

\usepackage{amsmath}

\newcommand{\2}{$_2$}

\submitjournal{PSJ}

\shorttitle{}
\shortauthors{}

\begin{document}

\title{Multiple moist climate equilibrium states on arid rocky M-dwarf planets:\\ A last-saturation tracer analysis}

\correspondingauthor{Feng Ding}
\email{fengding@g.harvard.edu}

\author[0000-0001-7758-4110]{Feng Ding}
\affiliation{School of Engineering and Applied Sciences, Harvard University, Cambridge, MA 02138, USA}

\author[0000-0003-1127-8334]{Robin D. Wordsworth}
\affiliation{School of Engineering and Applied Sciences, Harvard University, Cambridge, MA 02138, USA}
\affiliation{Department of Earth and Planetary Sciences, Harvard University, Cambridge, MA 02138, USA}



\begin{abstract}
Terrestrial-type exoplanets orbiting nearby red dwarf stars (M dwarfs) are the first potentially habitable exoplanets suitable for atmospheric characterization in the near future. Understanding the stability of water in cold-trap regions on such planets is critical because it directly impacts transmission spectroscopy observations, the global energy budget, and long-term surface water evolution. Here we diagnose the humidity distribution in idealized general circulation model (GCM) simulations of terrestrial-type exoplanets. We use the `tracer of last saturation' technique to study the saturation statistics of air parcels. We find that on synchronously rotating planets, the water vapor abundance in the nightside upper troposphere depends weakly on planetary rotation, while more water vapor builds up in the nightside lower troposphere on fast rotating planets. We then discuss how last-saturation statistics can elucidate the multiple moist climate equilibrium states on synchronously and asynchronously rotating arid planets. We show that the multiple moist climate states arise from the cold-trapping competition between the substellar upper atmosphere and cold surface regions. We find that fast synchronously rotating planets tend to trap surface water on the nightside as a result of their weak atmospheric and strong surface cold traps compared to the slow rotating case. These results elucidate the nature of the water cycle on arid rocky exoplanets and will aid interpretation of atmospheric observations in the future.

\end{abstract}



\section{Introduction} \label{sec:intro}
As hundreds of Earth-sized planets are discovered beyond our solar system\footnote{\url{https://exoplanetarchive.ipac.caltech.edu/}}, the new frontier of exoplanet science will soon be characterization of atmospheres on these terrestrial planets. For the next few years, the most observable of these planets will be those orbiting in the habitable zone around nearby M-dwarfs. Potentially habitable exoplanets around M-dwarfs may be very different from Earth because they are close to their host stars and expected to be tidally resonant or locked due to strong gravitational tidal interactions. In the most extreme case, close-in planets in circular orbits will be synchronously rotating, with a permanent dayside and nightside \citep{Kasting1993habitable, barnes2017tidal}.

M-dwarf stars have a prolonged pre-main-sequence phase \citep{baraffe2015pre}, which has been predicted to cause extensive early water loss on any rocky exoplanets orbiting in their habitable zones \citep{ramirez2014premain, luger2015waterloss, tian2015waterloss}. As a result, many such planets may be water-poor. 
Understanding the potential atmospheric water content of such planets is vital for several reasons. 
First, for transit observations, water vapor profiles near the terminator directly impact the transmission spectroscopy of water vapor. Second, water vapor is critical to the energy budget of the climate system: for example,  sub-saturated atmospheric regions can behave like ``radiator fins'' and stabilize the climate \citep{yang2014twocolumn,ding2020waterrich}, similar to what happens in the subtropics on the Earth \citep{pierrehumbert_thermostats_1995}. 
Third, the atmospheric water vapor concentration above the surface water reservoir determines whether surface condensation or evaporation should occur, which controls the long-term evolution of the surface water inventory. For example, \citet{ding2020arid} showed that on arid tidally locked planets covered by ice caps on the nightside surface, the difference between nightside atmospheric water vapor concentration and the saturation value at the surface determines whether the surface water will experience sublimation or condensation and hence whether the remaining water inventory will migrate towards the substellar region, forming an `oasis', or still be stably trapped on the nightside surface.
While \citet{ding2020arid} only focused on slowly rotating planets, here we conduct GCM simulations to explore multiple moist climate equilibrium states in a broader parameter space. 

Explaining the emergence of multiple moist climate systems requires a full understanding of water vapor distribution in the atmosphere.
In a moist climate, the primary source and sink for atmospheric water vapor are evaporation of surface water and precipitation in the atmosphere, respectively. Beyond this, the atmospheric humidity distribution is strongly influenced by atmospheric dynamics. An idealized model of non-local large-scale control of tropospheric humidity, called the last-saturation model or advection–condensation model, has been proposed for understanding the humidity distribution in Earth's atmosphere \citep{pierrehumbert1998subtropical, pierrehumbert2007relative, pierrehumbert2016nondilute}. This model assumes the mixing ratio of water vapor in an air parcel is conserved along any segment of a trajectory unless it becomes saturated and loses water by precipitation. In general, there are two ways of describing advective-condensation processes: the Lagrangian approach and the Eulerian approach. For the Lagrangian approach, \citet{pierrehumbert1998subtropical} used the backward Lagrangian trajectory technique to reconstruct relative humidity in the Earth’s subtropics and showed close agreement with Meteosat satellite observations. \citet{yang2019a3dcomparison} also used the same method to interpret the differences in nightside humidity distributions on tidally locked planets among some GCM simulations. The back trajectory technique can reconstruct the humidity distribution by the three-dimensional (3D) wind and temperature fields with a high spatial resolution, but it only provides the humidity distribution as a single snapshot. For climate systems with strong transient motions, especially on tidally locked planets \citep{merlis2010tidally, pierrehumbert2019tidelocked}, this technique is not optimal for presenting last-saturation statistics in the equilibrium climate system averaged over a long time period. 

Here we instead take an Eulerian approach to describe the last-saturation statistics in GCM simulations with tracers of last saturation for the first time. These tracers of last-saturation are advected by the air flow in the same way as water vapor in the GCM, and their concentration intrinsically contain information regarding water vapor's last-saturation statistics over a long time period.
 \citet{galewsky2005last} first developed this method to study the  dryness of the Earth's subtropical troposphere \added{in both idealized GCM simulations and a reanalysis dataset}. They found that the subsaturation is primarily controlled by the isentropic transport by midlatitude eddies and that diabatic descending transport from the tropical upper troposphere plays a secondary role. However, this method has not been applied to tidally locked planets so far. 
 
We briefly introduce this method and our GCM setup in Section~\ref{sec:method}. We show the multiple moist climate equilibrium states on slowly-rotating and rapidly-rotating tidally locked arid planets in Section~\ref{sec:arid}.
We diagnose the humidity distributions on slowly-rotating and rapidly-rotating tidally locked planets in Section~\ref{sec:result}. We then discuss how the last-saturation statistics elucidates the multiple moist climate equilibrium states on arid planets in Section~\ref{sec:connection} and present our conclusions in Section~\ref{sec:conclusion}.


\section{Method} \label{sec:method}

\subsection{Idealized moist general circulation model}

Previously, to study the hydrological cycles on arid planets, we developed a moist GCM that calculates the radiative transfer in the line-by-line approach and uses the Manabe moist convective adjustment scheme in \citet{ding2020arid}. Here, to study a wider range of climate states in an efficient way, we choose instead to simplify the radiation and convection calculations and make the moist GCM more idealized. 

\subsubsection{Two-stream gray radiation scheme}
Here we use a gray-gas radiation scheme, which is a standard approach for exploring climate states over a wide range for Earth-like atmospheres \citep{ogorman2008water} and for atmospheres on terrestrial exoplanets \citep{merlis2010tidally,koll2015phasecurve,koll2016heat}.
Following \citet{merlis2010tidally}, the longwave optical depth $\tau$ in our model has two contributing factors

\begin{equation} \label{eq:tau}
\tau(p) = \kappa_{wv} \int_{0}^{p} q(p') \frac{dp'}{g} + \tau_{dry} \frac{p}{p_0}.
\end{equation}
Here $\kappa_{wv}$ and $q$ are the absorption cross-section and the mass concentration of water vapor, $g$ is surface gravity, $p_0$ is the global mean surface pressure and $\tau_{dry}$ is the total optical thickness of the non-condensible absorber, which is assumed to be well mixed in the atmosphere. In \citet{merlis2010tidally}, $\kappa_{wv} = 0.1\,\mathrm{m^2\,kg^{-1}}$ and $\tau_{dry} = 1.2$. Our model uses the same value for $\kappa_{wv}$. Note that in this approach, the angle of propagation is implicitly incorporated into Eq.~\eqref{eq:tau}.

\subsubsection{Convective parameterization}
Following \citet{galewsky2005last}, our idealized GCM only considers the large-scale condensation of water vapor. In any super-saturated grid box, the excess moisture is removed immediately. 
\citet{frierson2007convection} studied the effect of various convection schemes on Earth's zonally averaged climate: the large-scale condensation (LSC) only scheme used here, the Manabe moist convective adjustment (MCA) scheme used in \citet{ding2020arid}, and the simplified Betts-Miller (SBM) scheme used in \citet{merlis2010tidally}. The global surface temperature distributions are similar in simulations with those convection  schemes, while tropical precipitation is less in the SBM simulation than in the LSC and MCA simulations. Use of the simple LSC convection scheme allows us to focus entirely on the transport role of large-scale dynamics, which is the main aim of this study.

\subsection{GCM setup for arid planet simulations}
The other aspects of our model setup are similar, but not identical, to those used in \citet{merlis2010tidally}. The planet has the same radius and surface gravity as Earth’s and both the eccentricity and obliquity are zero. The surface of the planet is flat and the incoming stellar radiation above the substellar point is 1367 W m$^{-2}$. The atmosphere is made of N\2 and condensible water vapor. \deleted{Two planetary rotation periods are investigated: 50 Earth days for a tidally locked planet orbiting in the habitable zone around {a bright M star}, and 10 Earth days around {a dim M star (e.g., TRAPPIST-1).} }

\added{We assume the simulated planets are in synchronous rotation, which implies the rotation period equals the orbital period. We calculate the orbital period $P$ self-consistently using Kepler's third law \citep{wordsworth2015heat, kopparapu2016innerhz, haqq2018circulation} as
\begin{equation} 
P = 365 \mbox{ days} \left(\frac{L}{L_\odot}\right)^{3/4}\left(\frac{M}{M_\odot}\right)^{-1/2}
\end{equation}  
where $L/L_\odot$ is the luminosity of the host star scaled by the luminosity of the Sun, and $M/M_\odot$ is the mass of the host star in solar mass unit. Two planetary rotation periods are investigated here: 50 Earth days for a synchronously rotating terrestrial  planet orbiting in the habitable zone around a  M1V type star, and 10 Earth days around a M5V type star. }

To explore the multiple moist climate equilibrium states on arid planets first studied in  \citet{ding2020arid}, we perform an additional set of simulations with our idealized moist GCM. For each simulation, we first place the initial surface water inventory within 40$^\circ$ around the substellar point, and run the GCM for 1000 days.  If no condensation occurs on the nightside surface, we refer to this climate state as the substellar oasis state. Next, we place the initial surface water inventory on the nightside, and run the GCM with the same parameters for 1000 days. If no condensation occurs at the substellar tropopause, we refer to this climate state as the nightside icecap state. While \citet{ding2020arid} only reported GCM simulations on a slowly rotating planet with two different CO\2 concentrations, here 
we explore a broader parameter space using the idealized moist GCM. We run these simulations with various surface pressures ($p_0$ in Eq.~\ref{eq:tau}), optical thicknesses of  non-condensible components ($\tau_{dry}$ in Eq.~\ref{eq:tau}) and planetary rotation rates to verify the surface water state (parameters listed in Table~\ref{tab:surfwater}).

\begin{table}[htp] 
\caption{Parameters used in GCM simulations to verify the surface water state.}
\begin{center}
\begin{tabular}{lc}
\hline
\hline
Parameter & Values \\
\hline
Global mean surface pressure [bar] & 0.5, 1, 2, 3, 5 \\
Optical thickness of  non-condensible components  & 0, 0.5, 1, 2, 3 \\
Planetary rotation period [Earth day] & 10, 50 \\
\hline
\end{tabular}
\end{center}
\label{tab:surfwater}
\end{table}%

\subsection{Last saturation tracer calculation}

For the last saturation tracer calculations, the global model domain is divided into $N$ axisymmetric subdomains along the axis from the substellar point to the antistellar point, $D_i\ (i=1,\ldots,N)$. In other words, these $N$ subdomains are zonally symmetric in the tidally locked coordinate in which the tidally locked latitude of the substellar point and antistellar point are usually defined as $90^\circ$ and $-90^\circ$, respectively (\citealt{koll2015phasecurve}, Appendix B).

Each subdomain $D_i$ is associated with a unique tracer of last saturation, $T_i$. If a  grid box reaches water vapor  saturation in one subdomain, $D_i$, we set the concentration of tracer associated with $D_i$ to unity and all the remaining tracers to zero in that saturated grid box,

\begin{equation}
\mathcal{T}_i = 1, \qquad \mathcal{T}_j = 0\ (j=1,\ldots,N, \mathrm{but}\ j \neq i).
\end{equation}
This is performed whenever a grid box reaches saturation in the GCM. Otherwise, all the tracers are simply advected by the large-scale circulation, in the same way as water vapor. 

In general, an unsaturated grid box in the free atmosphere above the planetary boundary layer consists of air parcels last saturated in multiple subdomains. Since each subdomain is associated with its own tracer, the value of an individual tracer $\mathcal{T}_i$ in that grid box can be interpreted as the probability that the air parcel was last saturated in its corresponding subdomain. Then the time mean and zonal mean water vapor distribution in the tidally locked coordinate can be approximately reconstructed by the time and zonal mean probability of last saturation in each subdomain $D_i$ and the saturation concentrations of water vapor $q_{sat,i}$ determined by the average temperature in the respective subdomains,

\begin{equation} \label{eq:construct_q}
q(\mathbf{r}) \approx \sum_{i=1}^{N} \mathcal{T}_i (\mathbf{r}) \cdot q_{sat,i}.
\end{equation}
where $\mathbf{r}$ is the position of the grid box.
The two terms in the product of Eq.~\ref{eq:construct_q}, the concentration of last saturation tracer and the saturation concentration of water vapor, represent the effect of atmospheric circulation (i.e., advection) and condensation on the water vapor buildup in the grid box. 

\subsection{GCM setup for last-saturation tracer analysis}
The main purpose of our analysis with tracers of last saturation is to investigate the key physical processes for water vapor buildup on tidally locked planets, and how the humidity distribution is affected by planetary rotation. For this part of the analysis, we therefore assume a global ocean surface or aqua-planet, which is a common approach in GCM simulations of habitable tidally locked planets \citep{merlis2010tidally,wordsworth2011gj581d,yang2013innerhz,noda2017rotation,haqq2018circulation,ding2020waterrich}. Our diagnostics can be directly applied to those previous simulations. In addition, the basic analysis is equally applicable to arid planets with a substellar water region (within 40$^\circ$ around the substellar point, see Appendix~\ref{sec:qarid}).

We use the same surface pressure and optical thickness of the non-condensible components as in \citet{merlis2010tidally} ($p_0 = 1$ bar and $\tau_{dry} =1.2$) and the two planetary rotation periods in the arid planet simulations (50 and 10 Earth days).
In \citet{haqq2018circulation}, these two \replaced{type}{types} of planets are defined as ``slow rotators'' and ``Rhines rotators'' due to the change of the circulation regimes. For simplicity,  our aqua-planet simulations with rotation period of 50 days and 10 days are referred to as the `slow rotating simulation' and `fast rotating simulation', respectively. Both simulations are run for 3000 days (Earth day, irrespective of the planetary rotation rate). Averaged results over the last 600 days are presented.

As a validation, we have ensured  that our idealized moist GCM can reproduce the surface conditions of the slow rotating simulation in \citet{merlis2010tidally} (e.g., surface temperature in their Figure~1 and surface winds in their Figure~4), although our numerical method that solves the fluid dynamic equations and the moist convection representation mentioned above are different. This similarity arises from the fact that it is the radiative process that drives the atmospheric circulation and we choose a very similar radiation scheme.

When applying the last-saturation tracer analysis in our slow rotating and fast rotating simulations, the subdomains associated its last-saturation tracers are defined as follows: each domain spans $180^\circ/8=22.5^\circ$ of latitude in the tidally locked coordinate and $100$\,hPa of pressure thickness from pressure level of $100$\,hPa to $950$\,hPa except that the pressure thicknesses of the bottom subdomains are $250$\,hPa (so $N = 8\times7=56$ in our simulations and all the 56 subdomains are highlighted by the red boxes in the 56 subpanels of Figure~\ref{fig:tracer}).
 
\section{Multiple moist climate states on arid synchronously rotating planets} \label{sec:arid}

\begin{figure}[ht]
  \centering
  \includegraphics[width=\columnwidth]{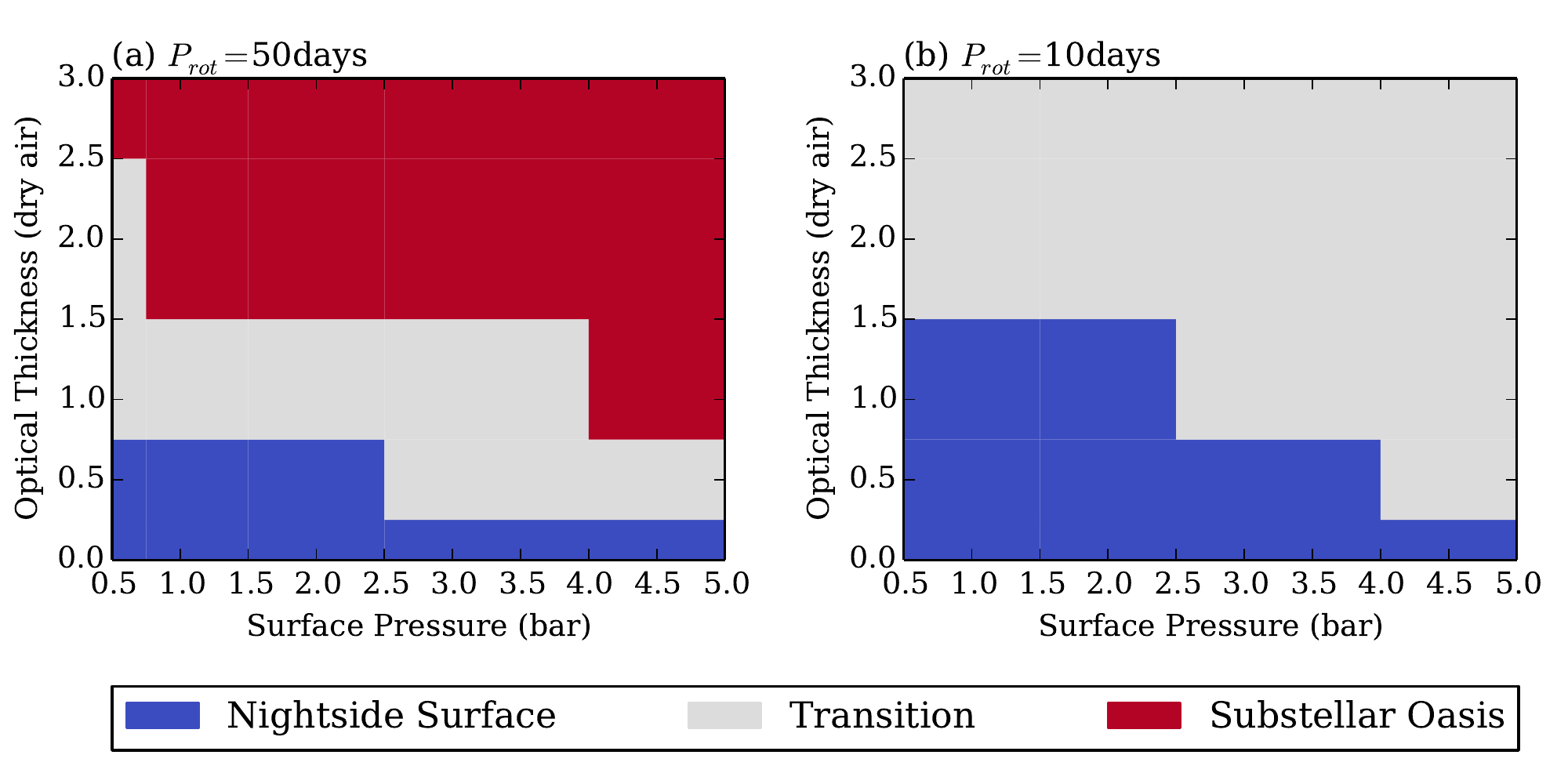}
  \caption{The states of surface water inventory on arid synchronously rotating planets as a function of surface pressure and total optical thickness of non-condensible components with planetary rotation period of 50 days (a) and 10 days (b). Red, gray and blue colors represent the substellar oasis state, the transition regime,  and the nightside icecap state.   }\label{fig:surfwater}
\end{figure}

First, we present our moist climate state results in the arid case as a function of rotation rate, surface pressure and atmospheric optical thickness. Figure~\ref{fig:surfwater} shows the states of surface water inventory on arid synchronously rotating planets under surface pressures ranging from 0.5 bar to 5 bar and optical thicknesses of non-condensible components ranging from 0 to 3.
The surface water \added{on slow rotators} is trapped on the nightside  when the atmospheric greenhouse effect is weak \added{and in the substellar region when the greenhouse effect is strong}, which is consistent with GCM simulation results with realistic radiative transfer calculation  in \citet{ding2020arid}. In addition, Figure~\ref{fig:surfwater} indicates that a thicker atmosphere tends to trap surface water on the dayside more effectively \added{on slow rotators}, because stronger upwelling motion at the the substellar tropopause strengthens the cold-trapping effect there.  Comparing Figure~\ref{fig:surfwater}a and b indicates that planetary rotation has a big impact on the moist climate states. Specifically, under the same conditions, fast rotating planets trap the surface water inventory on the nightside as ice more effectively than slow rotating planets. \added{In the parameter space we explore here, fast rotators have no substellar oasis state.}

In addition to the substellar oasis and nightside icecap states discussed in \citet{ding2020arid}, we find another climate state when the surface ice is stably trapped on the nightside but water vapor condensation occurs near the substellar tropopause. Here we refer to this state as the transition regime. \added{Table~\ref{tab:surfwater_characteristics} summarizes the key characteristics of  the three surface water states.}

\begin{table}[htp] 
\caption{Key characteristics of  planetary climates in the three surface water states.}
\begin{center}
\begin{splittabular}{lccBc}
\hline
\hline
Surface water state & Stable substellar surface water & Stable nightside surface water & Condensation in the upper atmosphere\\
\hline
Substellar oasis state & Yes & No & Yes \\
Transition regime & -- & Yes & Yes \\
Nightside icecap state & No & Yes & No \\
\hline
\end{splittabular}
\end{center}
\label{tab:surfwater_characteristics}
\end{table}%

In the next section, we discuss the last-saturation statistics on synchronously rotating aqua-planets and how the statistics is affected by planetary rotation. Following that, we use the last-saturation analysis to explain the emergence of multiple climate states on arid planets and why the equilibrium states are affected by planetary rotation.

\section{Humidity diagnostics using tracers of last saturation} \label{sec:result}

\begin{figure}[ht]
  \centering
  \includegraphics[width=\columnwidth]{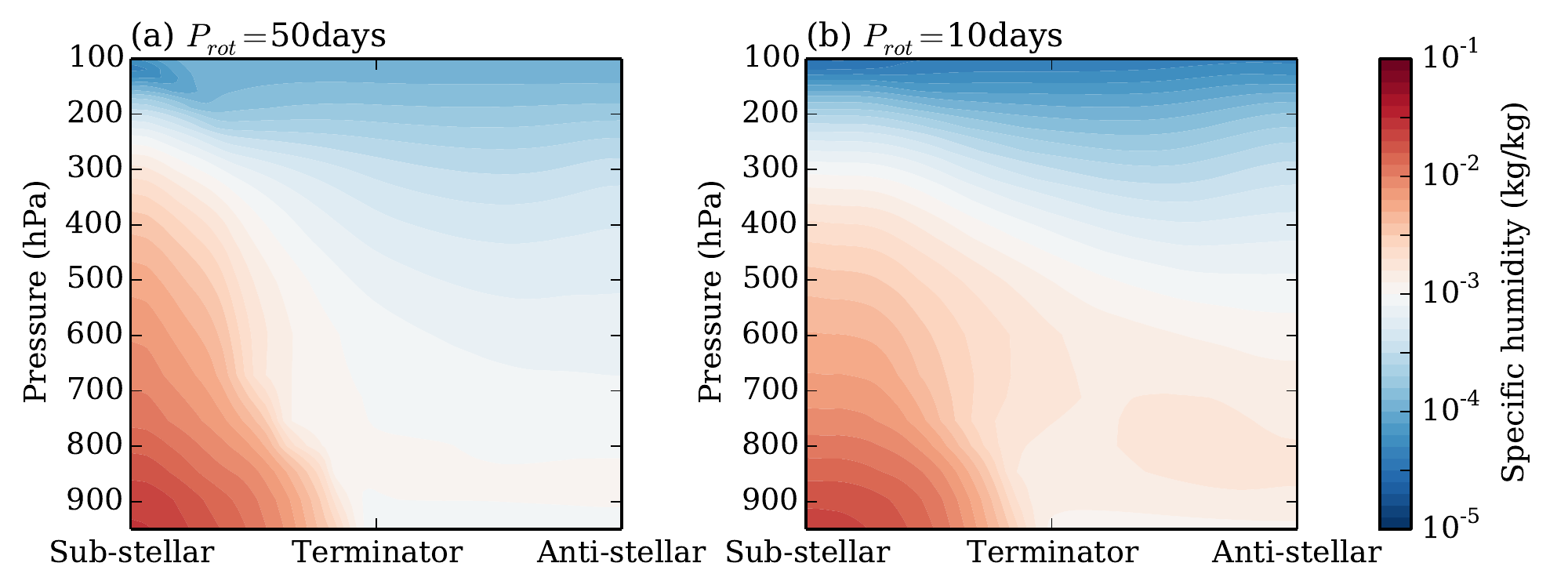}
  \caption{The time mean and zonal mean mass concentration of water vapor  calculated from the GCM in the tidally locked coordinate for the slow rotating simulation (a) and fast rotating simulation (b). }\label{fig:comp_q}
\end{figure}

To motivate the last saturation analysis further,
 the time mean and zonal mean water vapor concentration calculated from the GCM in the tidally locked coordinate for both the slow rotating and fast rotating aquaplanet simulations are shown in Figure~\ref{fig:comp_q}. There are two important features in the nightside distribution of water vapor. First, in both simulations, the water vapor concentration on the nightside decreases with height. Second, comparison between the two distributions shows that more water vapor builds up in the nightside lower atmosphere ($p>700$\,hPa) in the fast rotating simulation, but water vapor concentrations are similar in the upper troposphere in both simulations. As discussed in Section~\ref{sec:intro}, the water vapor abundance  in the nightside lower atmosphere  is important for the behavior of the nightside ``radiator fin'' on synchronously rotating planets and the long-term surface evolution on arid planets with limited surface water inventories. In this section, we diagnose the water vapor distribution using tracers of last saturation to explain the above features.

\subsection{Slow rotating simulation} 

\begin{figure}[ht]
  \centering
  \includegraphics[width=\columnwidth]{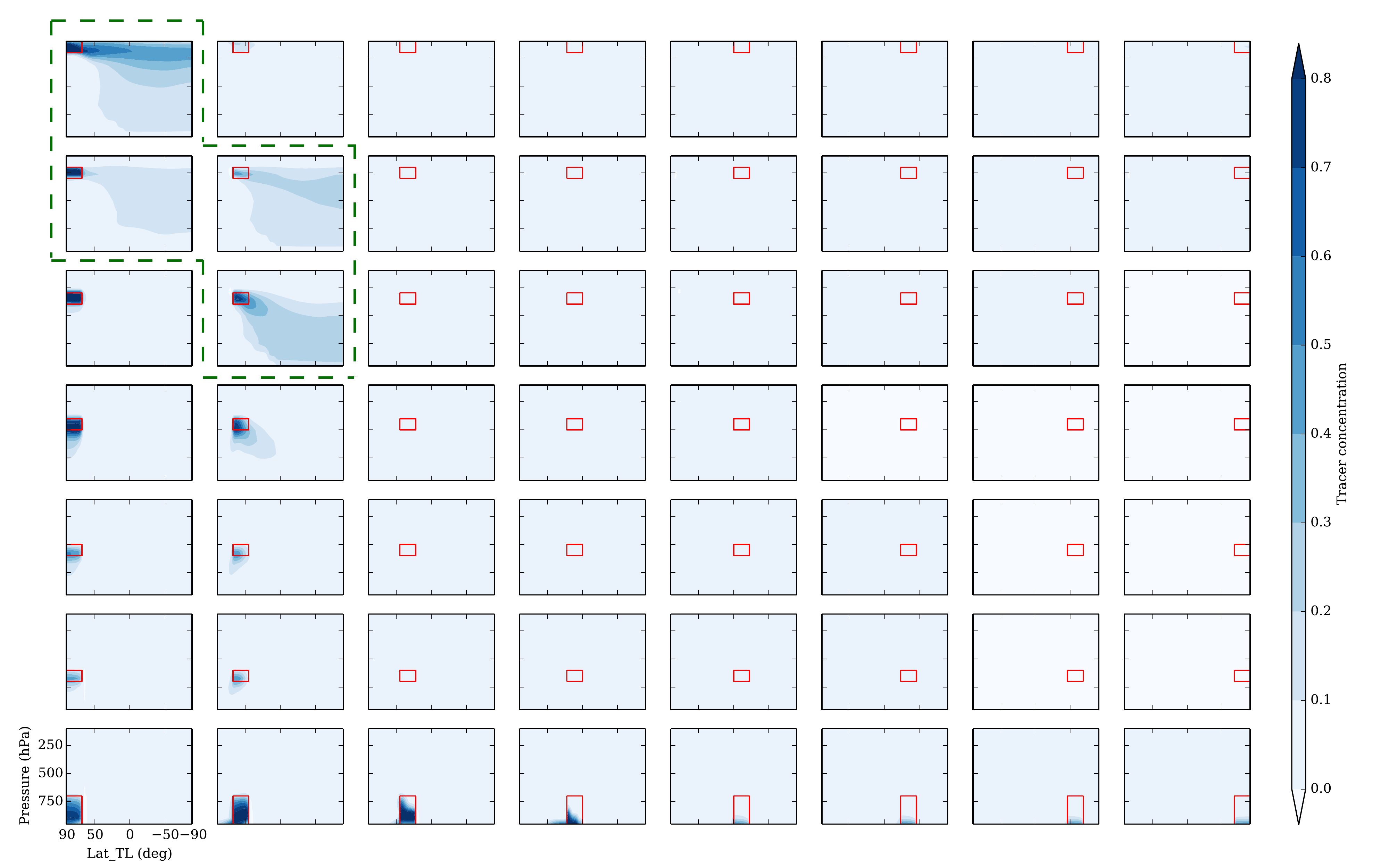}
  \caption{The time mean and zonal mean concentration of the 56 tracers of last saturation $\mathcal{T}_i$ in the tidally locked coordinate for the slow rotating simulation. Horizontal axes are  tidally locked latitude from the substellar point ($90^\circ$) to the antistellar point ($-90^\circ$); vertical axes are pressure (hPa). The 56 subdomains used in the calculation are highlighted by the red rectangles in each panel. The four subdomains enclosed by green dashed lines are the key cold-trapping region for water vapor buildup on the nightside. }\label{fig:tracer}
\end{figure}

Figure~\ref{fig:tracer} shows the time mean and zonal mean concentration of the 56 tracers of last saturation in the tidally locked coordinate for the slow rotating simulation. The value of each individual tracer can be interpreted as the probability that the air parcel was last saturated in its corresponding subdomain averaged over a long time period. The 56 subdomains used in the calculation are marked by the red rectangles in the 56 subpanels of Figure~\ref{fig:tracer}.

\begin{figure}[ht]
  \centering
  \includegraphics[width=\columnwidth]{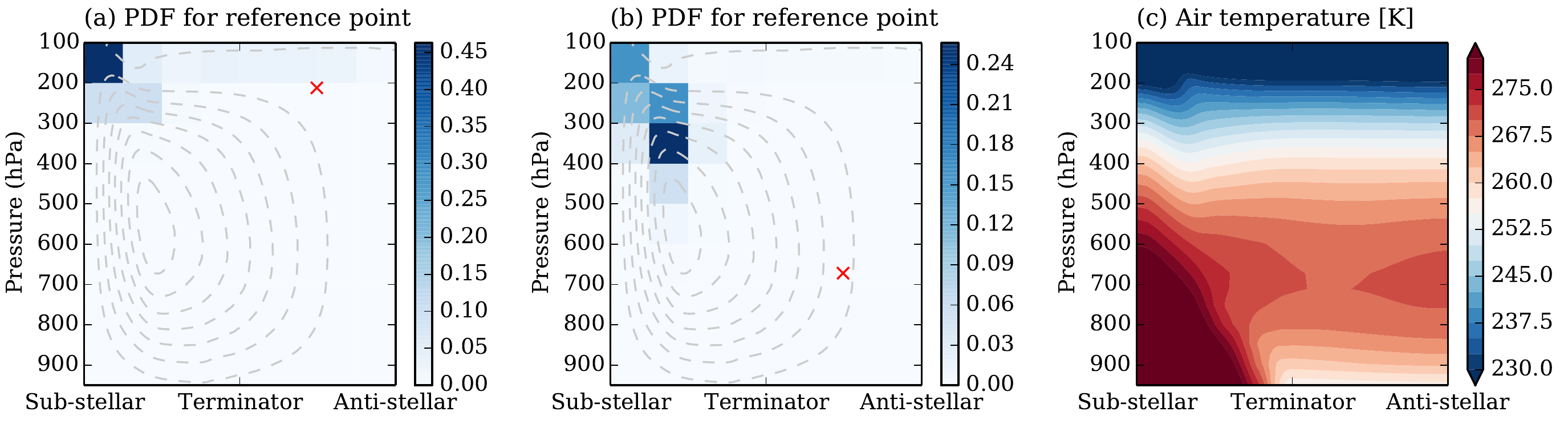}
  \caption{(a) Probability distribution function for locations of last saturation at the reference point (marked by the red cross in the upper troposphere on the nightside) in the tidally locked coordinate for the slow rotating simulation. The colors of the 56 subdomains are the values of the respective subdomain’s tracer concentration $\mathcal{T}_i$ at the reference point marked by the red cross, and also represent the chance that the air parcel at the reference point was last saturated in the respective subdomain. (b) Same as panel a, but the reference point is in the lower troposphere on the nightside. In both panel a and b, the time mean and zonal mean mass stream functions are marked by the dashed contours. (c) The time mean and zonal mean air temperature in the tidally locked coordinate for the slow rotating simulation.}\label{fig:pdf_slow}
\end{figure}

The distributions of last saturation tracers illustrate a straightforward interpretation of water vapor transport by the large-scale circulation in the slow rotating simulation. A thermally direct overturning circulation ascends within 40$^\circ$ around the substellar point and descends slowly in the remaining region (see dashed contours in Figure~\ref{fig:pdf_slow}a and b). In the atmosphere, water vapor condensation mainly occurs inside the upwelling branch of the overturning cell due to adiabatic cooling. 
In Figure~\ref{fig:tracer}, the tracers whose associated subdomains are inside the upwelling branch have large values in their respective subdomains and nearly vanish elsewhere, except in the upper level where the flow turns its direction and becomes horizontal. The tracer fields from the upper troposphere have distinguished sloping plumes of nonzero tracer concentration that descend towards the nightside. This key region acts as the cold trap of water vapor transport to the nightside and hence determines the water vapor abundance there. The most important cold-trapping regions are the subdomains enclosed by the green dashed lines in Figure~\ref{fig:tracer}.

The probability distributions of   last-saturation locations at reference points in the nightside troposphere (Figure~\ref{fig:pdf_slow}a and b) further illustrate this point and show more details of water vapor buildup on the nightside. 
The colors of the 56 subdomains in Figure~\ref{fig:pdf_slow}a and b represent the chance that the air parcel at the reference point was last saturated in the respective subdomain. For water vapor in the upper troposphere on the nightside, the key cold trap is the substellar tropopause (Figure~\ref{fig:pdf_slow}a). But for water vapor in the lower atmosphere, other than the substellar tropopause, the air parcel has nearly the same chance of last being saturated in the region right above the core of the overturning circulation (Figure~\ref{fig:pdf_slow}b), where both air temperature and the saturation concentration of water vapor are higher than at the tropopause. As a result, on the nightside of slowly and synchronously rotating planets,  there is a higher water vapor abundance  in the lower atmosphere  than in the upper atmosphere. 

\begin{figure}[ht]
  \centering
  \includegraphics[width=0.4\columnwidth]{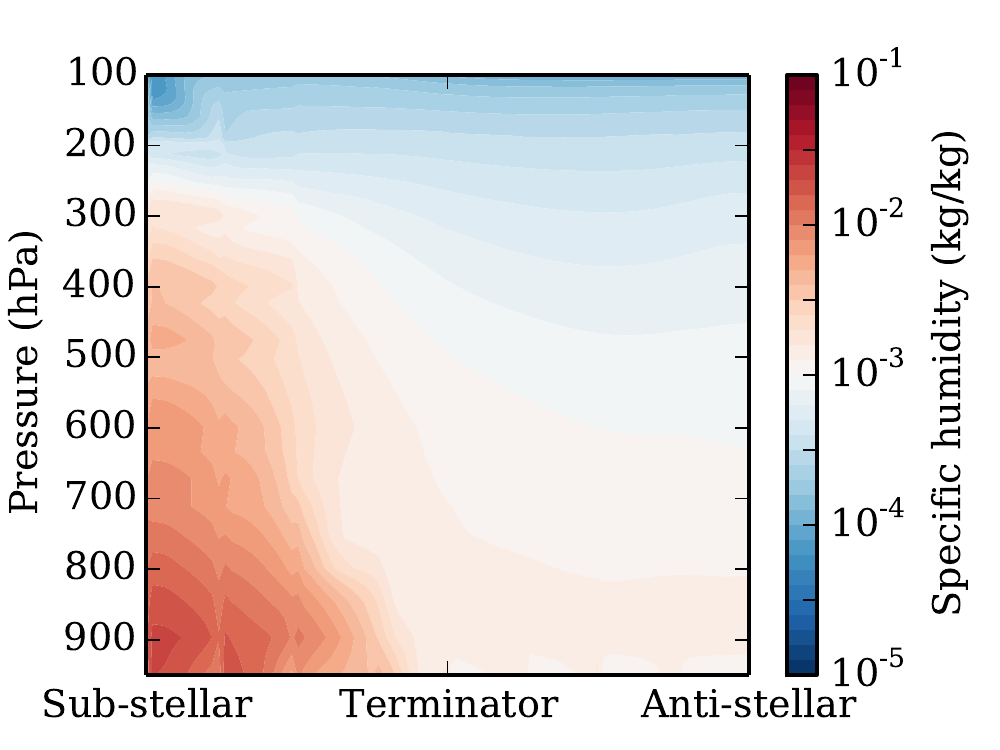}
  \caption{The time mean and zonal mean mass concentration of water vapor reconstructed from tracers of last saturation following Eq.~\ref{eq:construct_q} for the slow rotating simulation.}\label{fig:reconstruct_slow}
\end{figure}

In Eq.~\ref{eq:construct_q}, the water vapor concentration at the reference point can be interpreted as the summation of the product of the probability of last saturation in each subdomain and the saturation concentration of water vapor in the respective subdomain. Because of this, we can reconstruct the water vapor concentration in the atmosphere only using tracers of last saturation and air temperature in each subdomain. The reconstructed water vapor field in Figure~\ref{fig:reconstruct_slow} is very similar to the field calculated by the GCM (Figure~\ref{fig:comp_q}a), indicating our diagnostics with  tracers of last saturation captures essential physical processes for nightside water vapor buildup on slowly and synchronously rotating planets.


\subsection{Fast rotating simulation}

In the fast rotating simulation with planetary rotation period of 10 days, the atmospheric circulation is not zonally symmetric in the tidally locked coordinate. Although the meridional circulation is still dominated by the overturning circulation,   perturbations by stationary planetary waves are non-negligible, which was discussed in detail by applying the Helmholtz decomposition on atmospheric circulation in \cite{hammond2021rotational}. 

\begin{figure}[ht]
  \centering
  \includegraphics[width=0.9\columnwidth]{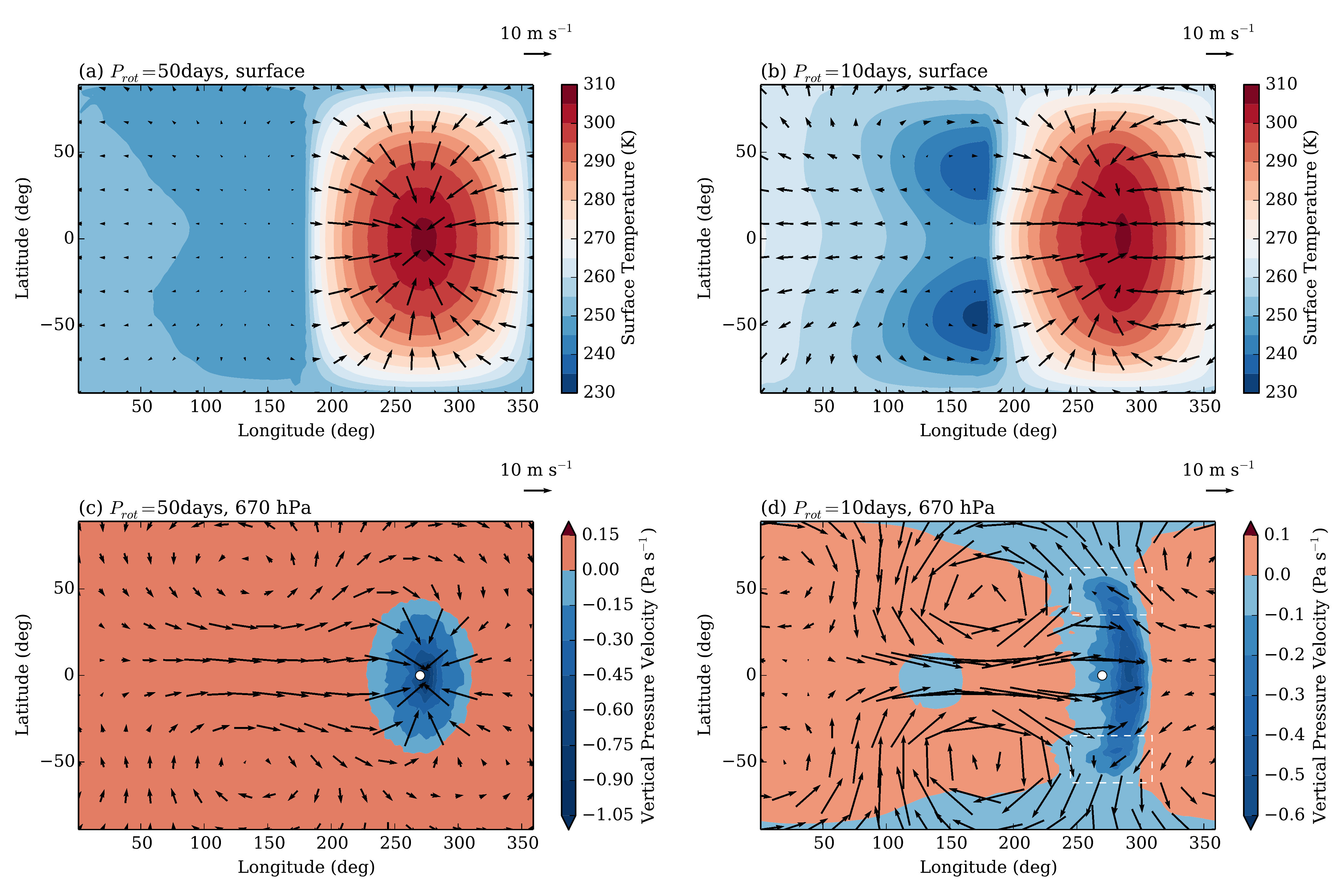}
  \caption{(a) Surface temperature (color contours) and surface wind (arrows) in the standard geographical coordinate in the slow rotating simulation. (b) Same as panel a, but for the fast rotating simulation. (c) Vertical pressure velocity (color contours) and horizontal wind (arrows) at 670 hPa in the slow rotating simulation. (d) Same as panel c, but for the fast rotating simulation. The longitude of the substellar point is 270$^\circ$ and marked by the white dot in panel c and d. In panel c and d, the dark blue region corresponds to the deep convection region around the substellar point. The white dashed lines in panel d highlight the key regions of water vapor transport to the nightside in the lower atmosphere of the fast rotating simulation. }\label{fig:ts_v}
\end{figure}

Figure~\ref{fig:ts_v}a and b compares the equilibrium surface state for both the slow and fast rotating simulations, revealing the change of large-scale dynamics.
Because of the modification by planetary wave perturbations, the hot spot of the surface temperature in the fast rotating simulation shifts eastwards from the substellar point (Figure~\ref{fig:ts_v}b). Meanwhile, two surface cold spots appear westwards of the west terminator associated with anticyclone-type circulations. \added{Figure~\ref{fig:ts_v}d shows super-rotating equatorial jet and mid-latitude stationary planetary wave patterns with the zonal wavenumber 1 in the lower atmosphere of the fast rotating simulation.} The interactions between the mean flow and planetary waves and the shifts of the temperature extrema were discussed in \added{both a linear shallow water model and an idealized dry GCM by } \citet{hammond2018wave}. \added{ \citet{hammond2018wave} showed during the spin-up stage of their GCM that the planetary wave pattern is Doppler-shifted eastwards to the equilibrium position as the zonal jet forms. The parameters of the default run in \citet{hammond2018wave} are identical to those of our fast rotating simulation (i.e., an Earth-sized planet with a 1~bar atmosphere dominated by N\2, the same incoming stellar radiation as on Earth and a 10 day orbital period) except that their atmosphere is dry with a longwave optical depth of 1. Their discussions on global circulation therefore qualitatively apply to our fast rotating simulation. }

\begin{figure}[ht]
  \centering
  \includegraphics[width=\columnwidth]{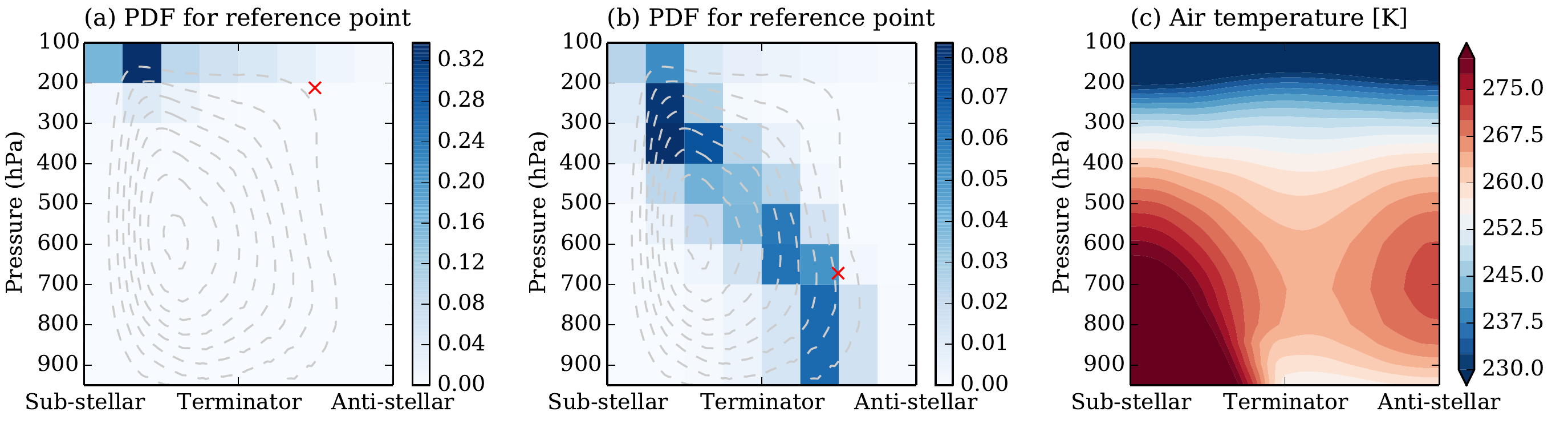}
  \caption{Same as Figure~\ref{fig:pdf_slow}, but for the fast rotating simulation with planetary rotation period of 10 days.}\label{fig:pdf_fast}
\end{figure}

The cold-trapping statistics of water vapor are greatly influenced by this change of atmospheric circulation compared to the slow rotating simulation. This becomes clear when we analyze the probability distributions for locations of last saturation at reference points in the nightside troposphere (Figure~\ref{fig:pdf_fast}a and b). In Figure~\ref{fig:pdf_fast}, the reference points are the same as in the slow rotating simulation (Figure~\ref{fig:pdf_slow}a and b). For the upper troposphere on the nightside, water vapor was still last saturated in the tropopause region (Figure~\ref{fig:pdf_fast}a), but shifted eastwards from the substellar point associated the eastward shift of surface hot spot in Figure~\ref{fig:ts_v}b. The tropopause temperatures are very close in the slow and fast rotating simulations, so that the water vapor concentration in the nightside upper troposphere are similar (Figure~\ref{fig:comp_q}). For the lower troposphere on the nightside, the last saturation statistics becomes more distributed compared to the slow rotating case.  Water vapor in the lower troposphere was still last saturated in the dayside upper atmosphere above the core of overturning cell, but the probability is only one third of that in the slow rotating simulation, as can be seen by comparing Figure~\ref{fig:pdf_slow}b and Figure~\ref{fig:pdf_fast}b. Figure~\ref{fig:pdf_fast}b shows that the water vapor in the lower troposphere was twice as likely to be last saturated in the middle to lower troposphere near the terminator, where the air is much warmer than in the upper troposphere in both simulations (Figure~\ref{fig:pdf_slow}c and Figure~\ref{fig:pdf_fast}c). Therefore, an excess of water vapor builds up in the nightside lower atmosphere on fast rotating planets. 

\added{Figure~\ref{fig:ts_v}d helps to elucidate the water vapor transport to the nightside in the lower atmosphere. Near both the northern and southern edges of the deep convection regions, the horizontal wind flows away from the deep convection region associated with the mid-latitude stationary planetary waves (specifically, the cyclonic gyres about 90$^\circ$ west of the deep convection region and the anticyclonic gyres about 90$^\circ$ east of the deep convection region). This large-scale air flow carries very humid air towards the nightside directly. It resembles the `atmospheric river' that brings sustained and heavy precipitation to the west coasts of North America in the Earth's climate system \citep{zhu1994atmriver,zhu1998atmriver}.  The time evolution of water vapor concentration at the same pressure level is consistent with this large-scale air flow north and south of the deep convection region in Figure~\ref{fig:ts_v}d (See Appendix~\ref{sec:qmovie} for details). In contrast, the edges of the deep convection region in the slow rotating simulation are all dominated by convergent flow towards the substellar point. This means that water vapor in the remaining subsidence region has to be advected from the upper atmosphere, as illustrated by the last-saturation statistics in Figure~\ref{fig:pdf_slow}b.} 

\added{These two types of global circulations have been investigated in previous studies of the circulation regimes of synchronously rotating terrestrial planets within the habitable zone. The circulation dominated by the global overturning circulation in our slow rotating simulation is analogous to the `slow rotation' circulation regime defined in \citet{haqq2018circulation} and the `Type-I' circulation regime defined in \citet{noda2017rotation}. The circulation with the insolation pattern relatively in phase with the deep convection region but out of phase with mid-latitude stationary waves of zonal wavenumber 1 is analogous to the `Rhines' regime defined in \citet{haqq2018circulation} and the `Type-II' circulation regime defined in \citet{noda2017rotation}. The `atmospheric river' phenomenon is a natural outcome of the misalignment of the deep convection region and mid-latitude stationary planetary waves on synchronously rotating terrestrial planets in such a transition circulation regime \citep{noda2017rotation, haqq2018circulation}.   }

\section{elucidation of multiple moist climate equilibrium states on arid planets} \label{sec:connection}

\subsection{Synchronously rotating arid planets}

\begin{figure}[ht]
  \centering
  \includegraphics[height=0.9\textheight]{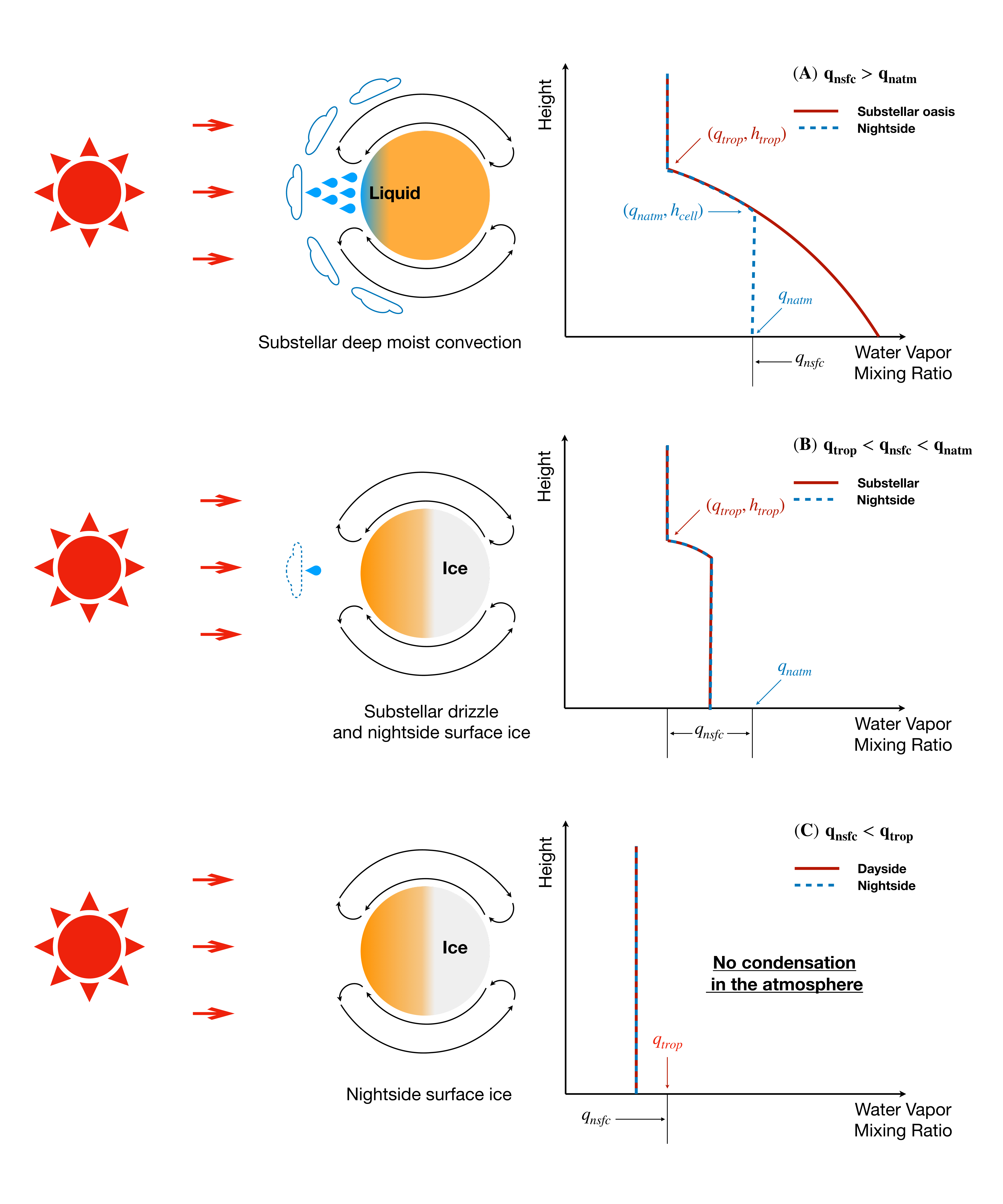}
  \caption{Schematic representation of the multiple moist climate equilibrium states and the respective moisture profiles on synchronously rotating planets, when the surface water is trapped by the substellar tropopause as a substellar oasis (a) and is trapped by the nightside surface as an icecap (c). Panel b is the transition regime. See Section~\ref{sec:connection} for the definition of mathematical symbols.}\label{fig:schematic}
\end{figure}

Having used last-saturation statistics to elucidate the behavior of atmospheric water vapor as a function of rotation on synchronously rotating planets, we now use these results to understand the multiple moist climate states presented in Section~\ref{sec:arid}. 
While we used the aqua-planet assumption in simulations with tracers of last saturation in Section~\ref{sec:result}, the basic analysis also applies to more arid situations. In the case where only substellar water is present, the water vapor profile in the nightside troposphere can be roughly separated into two regions by the altitude of the core of the overturning cell, $h_{cell}$. Above the altitude of $h_{cell}$, water vapor was mainly last saturated at the substellar tropopause (Figure~\ref{fig:pdf_slow}a and Figure~\ref{fig:pdf_fast}a). Let the tropopause height be $h_{trop}$ and the saturation water vapor concentration there be $q_{trop}$. Below the altitude of $h_{cell}$, water vapor was last saturated in warmer regions in the atmosphere (Figure~\ref{fig:pdf_slow}b and Figure~\ref{fig:pdf_fast}b). Assume the water vapor concentration there is roughly $q_{natm}$. On synchronously rotating planets with substellar surface water, $h_{trop}>h_{cell}, q_{trop}<q_{natm}$. In addition to the cold traps in the atmosphere, the nightside surface also behaves as an important cold trap where atmosphere water vapor may condense on the surface directly forming an ice layer. Let the saturation water vapor concentration at the nightside surface cold trap be $q_{nsfc}$. It is the comparison of the cold-trapping strength between the substellar atmosphere and the nightside surface (measured by $q_{trop}, q_{natm}, q_{nsfc}$)  that gives rise to the three moist climate regimes shown in Figure~\ref{fig:surfwater}.

\begin{itemize} 
\item{\it{Substellar oasis state.}}  When $q_{nsfc}>q_{natm}$, water vapor never condenses on the nightside surface anymore.  Precipitation forms within the upwelling branch of the overturning cell, leading to the formation of a substellar oasis surrounded by dry land (Figure~\ref{fig:schematic}a). This climate regime resembles Earth's tropical rain belt (Intertropical Convergence Zone, ITCZ) surrounded by the subtropical deserts and also Titan's polar methane lakes surrounded by tropical dunes \citep{mitchell2016titan}. Similar to on Earth and Titan, deep convective clouds form within the upwelling branch of the overturning cell. Some cloud particles will be carried by the upper-level divergent flow to the terminator, which may have consequences for transit observations \citep{komacek2020cloud, Pidhorodetska2020trappist1e, suissa2020dim}.   

\item{\it{Nightside icecap state.}} When $q_{nsfc}<q_{trop}$, the substellar tropopause loses its cold trap role. Any substellar water area will slowly migrate to the nightside and eventually be cold trapped as ice caps by the  freezing nightside  surface. In the equilibrium state,
there will be no condensation in the atmosphere and the dayside hemisphere is completely dry (Figure~\ref{fig:schematic}c). In Figure~\ref{fig:surfwater}, this state shows up when the nightside surface is cold because of the lack of insolation and weak greenhouse effect. 

\item{\it{Transition regime.}} The above two climate regimes are very similar to the the bistable moist climate states discussed for close-in arid exoplanets in \citet{leconte2013bistable}, except that \citet{leconte2013bistable} focused on planets receiving incoming radiation much higher than the runaway greenhouse threshold. 
\citet{ding2020arid}  discussed the above two cases for synchronously rotating planets in the habitable zone, and how they can be regulated by geological processes. However, there is in addition a third regime where $q_{trop}<q_{nsfc}<q_{natm}$ and the cold traps at both the nightside surface and substellar tropopause are effective (Figure~\ref{fig:schematic}b). Since $q_{nsfc}<q_{natm}$, the surface water inventory is cold trapped by the nightside surface, same as in the nightside icecap state. On the other hand, because $q_{nsfc}>q_{trop}$, tiny amount of condensation still occurs near the substellar tropopause, producing a thin layer of cirrus clouds. Whether the tiny amount of precipitation that these clouds produce can reach the substellar surface depends on the competition between the falling speed and evaporation of raindrops, both of which are sensitive to the raindrop size \citep{loftus2021rain}.  This question cannot be answered in large-scale GCMs, and should be studied in numerical models that both resolves convection scale motions and employs appropriate microphysical schemes. 
So this transition climate regime is more complicated than the regimes with a single cold trap. It bears similarities both with present-day Mars with polar water ice caps \citep{montmessin2017marswater}, and a cold early Mars scenario with both polar and highland ice \citep{wordsworth2013mars}. 

\end{itemize}

To sum up, it is the competition of cold-trapping between the substellar upper atmosphere and nightside surface that gives rise to the emergence of the multiple moist climate states on synchronously rotating arid planets seen in Figure~\ref{fig:surfwater}.  Now we can use this concept to explain why surface water tends to be trapped on the nightside on fast rotating planets.
First, $q_{natm}$ is slightly larger on fast rotating planets with the presence of substellar water (Figure~\ref{fig:comp_q}), which was elucidated by our tracer of last saturation analysis in Section~\ref{sec:result}. Second, the nightside surface cold spots  on fast rotating planets are much colder than the uniform nightside surface temperature on slow rotating ones (Figure~\ref{fig:ts_v}a and b), indicating a much smaller $q_{nsfc}$. Both larger $q_{natm}$ and smaller $q_{nsfc}$ on fast rotating planets make the surface water more likely to be trapped on the nightside as ice caps.

\added{In a real climate, many factors that are not included in our idealized GCM simulations could affect the  cold-trapping competition between the substellar upper atmosphere and nightside surface by influencing the tropopause temperature and nightside surface temperature. These factors include real-gas radiative transfer \citep{ding2020arid}, strong shortwave absorbers in the upper atmosphere such as ozone, methane and aerosols, radiative effects of clouds, turbulent mixing in the highly stratified thermal inversion layer above the nightside surface \citep{joshi2020boundary}. It will be interesting to explore these effects with more complex GCM simulations in the future.}

\subsection{Asynchronously rotating arid planets}
Arid planets with non 1:1 spin-orbit resonance will experience diurnal cycles, and will also have seasonal cycles if their obliquities are non-zero. Here we briefly discuss their likely multiple moist climate states. Although this topic is worth exploring in detail using 3D GCMs with more complex land models in the future, the basic cold-trapping principles for determining the equilibrium moist climate state described in the last section will still apply. 

When the surface cold trap (i.e., the coldest surface area in the long-term averaged climate) dominates the hydrological cycle, permanent ice caps will inevitably be drawn there. On arid planets with low obliquity, this will be the polar region; on planets with obliquity higher than 45$^\circ$, it moves to the equator (\citealt{pierrehumbert2010principle}, Chapter 7.3).
The former case resembles present-day Mars with polar water ice caps. Because of the lack of a massive moon and the proximity of Jupiter, the obliquity of Mars evolves chaotically and can drift to values as large as 47$^\circ$ \citep{laskar2002mars}.  
When Mars's obliquity is high, the surface ice migrates to the tropics -- a process which has resulted in the formation of glaciers on Mars in the recent past \citep{head2005tropical,forget2006formation}.

When the substellar tropopause cold trap instead dominates the hydrological cycle, surface water should appear under the upwelling branch of the thermally direct overturning circulation, which is usually the warmest place on the planet. On arid planets with obliquity less than 10$^\circ$, it is the tropics; on planets with  obliquity higher than 45$^\circ$, the warmest and coldest place over the course of an astronomical year are the summer and winter pole, respectively (\citealt{pierrehumbert2010principle}, Chapter 7.3). The latter case resembles present-day Titan, with methane replacing water as the working condensible component. For instance, the Imaging Science Subsystem (ISS) and the Visual and Infrared Mapping Spectrometer (VIMS) on Cassini observed rapid migration of convective methane clouds from the high latitudes of one hemisphere to the other during the solstice season. Cassini ISS also observed a large dune field near the equator, implying dry surface condition at low latitudes \citep{mitchell2016titan}. Recently, \citet{mackenzie2019titanlake} observed the disappearance of polar `phantom lakes' or shallow ponds in the winter season and interpreted this as a result of methane  evaporation and infiltration into a porous regolith. This climate regime on arid planets with strong seasonal cycles strongly resembles the substellar oasis state on synchronously rotating planets, with the `substellar oasis' oscillating between the two poles in this case. 

The transition regime subject to diurnal and seasonal cycles when both the atmospheric and surface cold traps are effective  becomes even more complex. But it is relevant to many interesting climate evolution topics and should be investigated in more detail in future studies. These topics include the ice distribution and \added{episodic warming on early Mars \citep{wordsworth2013mars,wordsworth2021episodic}}, and the delayed onset of runaway greenhouse on terrestrial planets only covered by polar surface water \citep{abe2011dry,kodama2018dry}. 


\section{Conclusions} \label{sec:conclusion}
We studied cold trapping in idealized GCM simulations of synchronously rotating exoplanets using tracers of last saturation and discussed the implications for long-term surface water evolution on arid planets. Our main findings are as follows:
\begin{enumerate}
\item For the nightside upper troposphere above the center of the overturning cell, water vapor is mainly cold trapped by the substellar tropopause, regardless of the planetary rotation rate. 

\item For the nightside lower troposphere, water vapor is mainly cold trapped by the region right above the center of the overturning cell on slow rotating planets. On fast rotating planets, the water vapor abundance in the lower atmosphere is higher than on slow rotating planets given the same condition, because the last saturation locations include the warm lower atmosphere near the terminator due to planetary wave perturbations.

\item On synchronously rotating arid planets with limited surface water inventories, there are multiple moist climate states. We explored the parameter space over a wide range of surface pressures, optical thicknesses of the non-condensible components, and planetary rotation rates. Our results show that fast rotating planets tend to trap the surface water inventory on the nightside surface as ice caps.

\item Multiple moist climate equilibrium states emerge from the competition between cold-trapping strength in the substellar upper atmosphere and on the coldest regions of the surface. This broad concept can be applied to both synchronously rotating and asynchronously rotating planets.

\end{enumerate}

\acknowledgments
We thank the two referees for thoughtful comments that improved the manuscript. 
We thank Daniel Koll for providing the Python scripts that convert GCM output from standard/Earth-like coordinates into a tidally locked coordinate system ({\url{https://github.com/ddbkoll/tidally-locked-coordinates}}). R.W. acknowledges funding support from NASA/VPL grant UWSC10439.
The GCM simulations in this paper were run on the FASRC Cannon cluster supported by the FAS Division of Science Research Computing Group at Harvard University. The idealized GCM with last-saturation tracers and outputs in the tidally locked coordinate are available on Github at \url{https://github.com/fdingdfdfdf/fmspcm_lstracer}. 



\appendix
\section{Humidity distribution on arid planets with substellar water} \label{sec:qarid}

\begin{figure}[ht]
  \centering
  \includegraphics[width=\columnwidth]{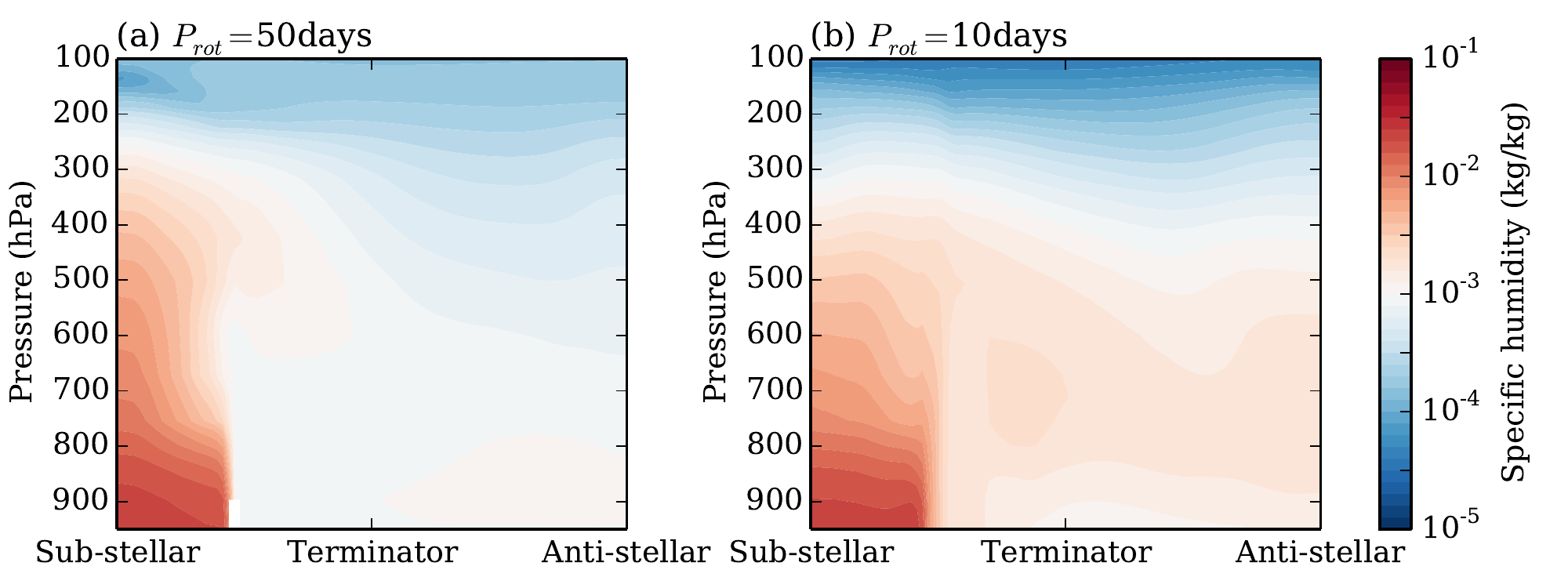}
 \caption{Same as Figure~\ref{fig:comp_q}, but the surface water only distributes within 40$^\circ$ around the substellar point in both simulations.}\label{fig:comp_qarid}
\end{figure}

To validate that our analysis with tracers of last saturation on aqua-planets can be equally applied to arid planets with substellar water, we conduct another two simulations. The GCM parameters are the same as in both aqua-planet runs, except that the surface water only distributes within 40$^\circ$ around the substellar point. Figure~\ref{fig:comp_qarid} shows the time and zonal mean states of water vapor distribution. Compared to the two aqua-planet runs, the lower atmosphere above the dayside dry land around the substellar water becomes much drier because of the lack of water vapor fluxes from the surface. However, the water vapor distribution elsewhere is very close to those in the two aqua-planet runs shown in Figure~\ref{fig:comp_q}. This indicates that the key physical processes controlling the nightside water vapor buildup in the arid planet simulations with substellar water is very similar to those in the aqua-planet simulations.

\begin{figure}[ht]
\begin{interactive}{animation}{qmovie.mp4}
  \centering
  \includegraphics[width=0.7\columnwidth]{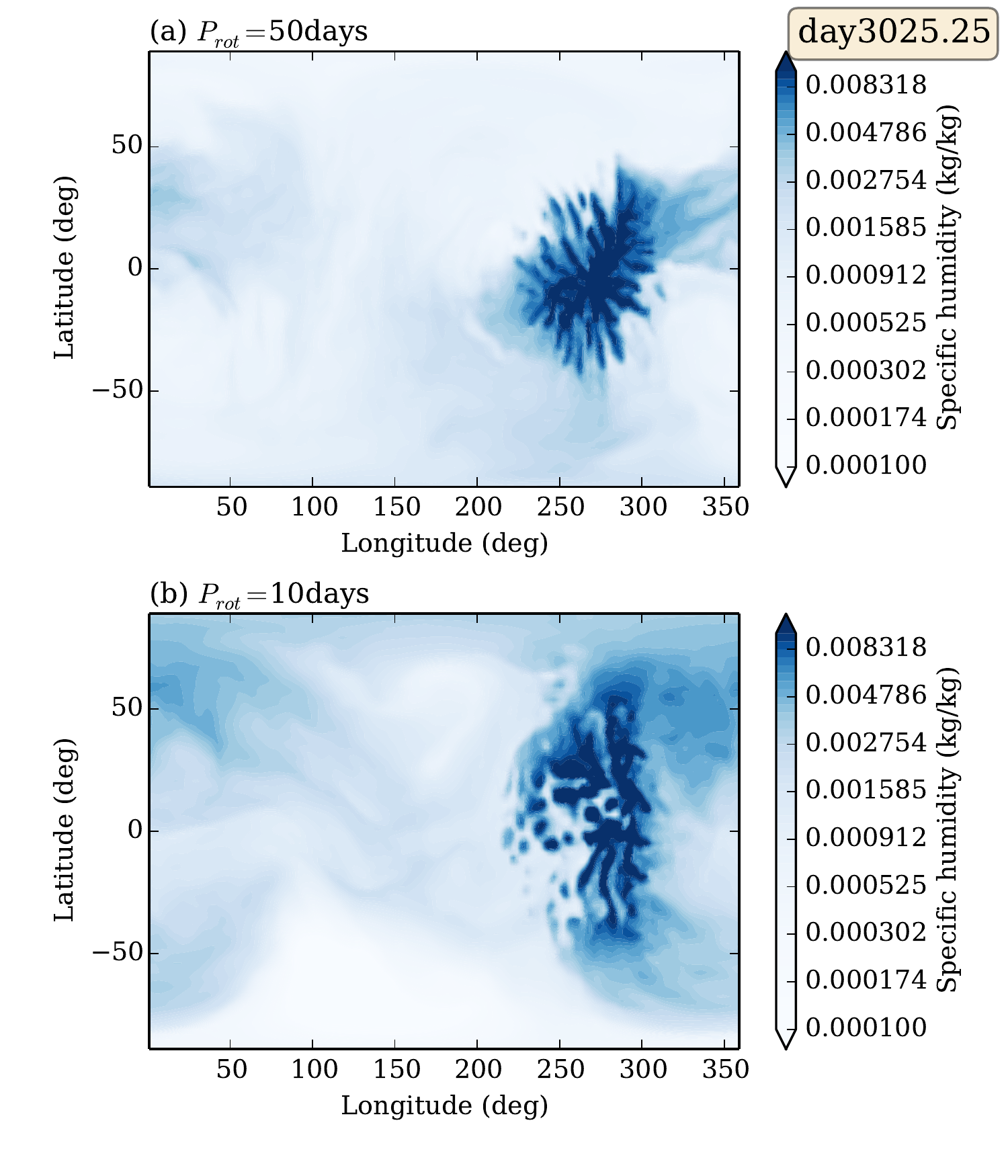}
\end{interactive}
 \caption{Time evolution of water vapor concentration at 670 hPa from the model day 3000 to day 3030 in the slow rotating simulation (a) and fast rotating simulation (b). The substellar point is 270$^\circ$.}\label{fig:qmovie}
\end{figure}

\added{\section{Time evolution of lower-atmosphere water vapor concentration in the aqua-planet GCM simulations}
\label{sec:qmovie}
To better understand the physical mechanism of water vapor transport to the nightside, we continue the aqua-planet simulations for another 30  days and use the 6-hourly snapshots to create an animation of water vapor distribution at 670 hPa (Figure~\ref{fig:qmovie}). In the slow rotating simulation, Figure~\ref{fig:qmovie}a shows the deep convection region is controlled by convergent flow towards the substellar point. So water vapor in the 
remaining subsidence region has to be advected in the upper atmosphere, which is consistent with the last saturation analysis in Figure~\ref{fig:pdf_slow}b. However, in the fast rotating simulation, water vapor in the lower atmosphere is carried by the horizontal wind flowing away from the deep convection region both north and south of the deep convective region, resulting in higher water vapor concentrations on the nightside. Other than stationary flows, this animation also present interesting transient wave activities in both simulations, especially on the nightside, which may be  responsible for the weak horizontal moisture gradient on the nightside (Figure~\ref{fig:comp_q}a and b). There are also granular structures in the specific humidity convection cells that are probably associated with grid-scale condensation (our idealized GCM has no sub-grid convection parameterization). Similar granular structures were also observed in the idealized GCM simulations that explored the effects of idealized convection schemes on Earth's tropical circulations \citep{frierson2007convection}. }
%
%

%

\bibliography{fmspcm}

\begin{thebibliography}{}
\expandafter\ifx\csname natexlab\endcsname\relax\def\natexlab#1{#1}\fi
\providecommand{\url}[1]{\href{#1}{#1}}
\providecommand{\dodoi}[1]{doi:~\href{http://doi.org/#1}{\nolinkurl{#1}}}
\providecommand{\doeprint}[1]{\href{http://ascl.net/#1}{\nolinkurl{http://ascl.net/#1}}}
\providecommand{\doarXiv}[1]{\href{https://arxiv.org/abs/#1}{\nolinkurl{https://arxiv.org/abs/#1}}}

\bibitem[{{Abe} {et~al.}(2011){Abe}, {Abe-Ouchi}, {Sleep}, \&
  {Zahnle}}]{abe2011dry}
{Abe}, Y., {Abe-Ouchi}, A., {Sleep}, N.~H., \& {Zahnle}, K.~J. 2011,
  Astrobiology, 11, 443, \dodoi{10.1089/ast.2010.0545}

\bibitem[{{Baraffe} {et~al.}(2015){Baraffe}, {Homeier}, {Allard}, \&
  {Chabrier}}]{baraffe2015pre}
{Baraffe}, I., {Homeier}, D., {Allard}, F., \& {Chabrier}, G. 2015, \aap, 577,
  A42, \dodoi{10.1051/0004-6361/201425481}

\bibitem[{{Barnes}(2017)}]{barnes2017tidal}
{Barnes}, R. 2017, Celestial Mechanics and Dynamical Astronomy, 129, 509,
  \dodoi{10.1007/s10569-017-9783-7}

\bibitem[{{Ding} \& {Pierrehumbert}(2020)}]{ding2020waterrich}
{Ding}, F., \& {Pierrehumbert}, R.~T. 2020, \apjl, 901, L33,
  \dodoi{10.3847/2041-8213/abb941}

\bibitem[{{Ding} \& {Wordsworth}(2020)}]{ding2020arid}
{Ding}, F., \& {Wordsworth}, R.~D. 2020, \apjl, 891, L18,
  \dodoi{10.3847/2041-8213/ab77d1}

\bibitem[{Forget {et~al.}(2006)Forget, Haberle, Montmessin, Levrard, \&
  Head}]{forget2006formation}
Forget, F., Haberle, R.~M., Montmessin, F., Levrard, B., \& Head, J. 2006,
  Science, 311, 368

\bibitem[{{Frierson}(2007)}]{frierson2007convection}
{Frierson}, D. M.~W. 2007, Journal of Atmospheric Sciences, 64, 1959,
  \dodoi{10.1175/JAS3935.1}

\bibitem[{{Galewsky} {et~al.}(2005){Galewsky}, {Sobel}, \&
  {Held}}]{galewsky2005last}
{Galewsky}, J., {Sobel}, A., \& {Held}, I. 2005, Journal of Atmospheric
  Sciences, 62, 3353, \dodoi{10.1175/JAS3533.1}

\bibitem[{{Hammond} \& {Lewis}(2021)}]{hammond2021rotational}
{Hammond}, M., \& {Lewis}, N.~T. 2021, arXiv e-prints, arXiv:2102.11760.
\newblock \doarXiv{2102.11760}

\bibitem[{{Hammond} \& {Pierrehumbert}(2018)}]{hammond2018wave}
{Hammond}, M., \& {Pierrehumbert}, R.~T. 2018, \apj, 869, 65,
  \dodoi{10.3847/1538-4357/aaec03}

\bibitem[{{Haqq-Misra} {et~al.}(2018){Haqq-Misra}, {Wolf}, {Joshi}, {Zhang}, \&
  {Kopparapu}}]{haqq2018circulation}
{Haqq-Misra}, J., {Wolf}, E.~T., {Joshi}, M., {Zhang}, X., \& {Kopparapu},
  R.~K. 2018, \apj, 852, 67, \dodoi{10.3847/1538-4357/aa9f1f}

\bibitem[{Head {et~al.}(2005)Head, Neukum, Jaumann, Hiesinger, Hauber, Carr,
  Masson, Foing, Hoffmann, Kreslavsky, {et~al.}}]{head2005tropical}
Head, J., Neukum, G., Jaumann, R., {et~al.} 2005, Nature, 434, 346

\bibitem[{{Joshi} {et~al.}(2020){Joshi}, {Elvidge}, {Wordsworth}, \&
  {Sergeev}}]{joshi2020boundary}
{Joshi}, M.~M., {Elvidge}, A.~D., {Wordsworth}, R., \& {Sergeev}, D. 2020,
  \apjl, 892, L33, \dodoi{10.3847/2041-8213/ab7fb3}

\bibitem[{{Kasting} {et~al.}(1993){Kasting}, {Whitmire}, \&
  {Reynolds}}]{Kasting1993habitable}
{Kasting}, J.~F., {Whitmire}, D.~P., \& {Reynolds}, R.~T. 1993, \icarus, 101,
  108, \dodoi{10.1006/icar.1993.1010}

\bibitem[{{Kodama} {et~al.}(2018){Kodama}, {Nitta}, {Genda}, {Takao}, {O'ishi},
  {Abe-Ouchi}, \& {Abe}}]{kodama2018dry}
{Kodama}, T., {Nitta}, A., {Genda}, H., {et~al.} 2018, Journal of Geophysical
  Research (Planets), 123, 559, \dodoi{10.1002/2017JE005383}

\bibitem[{{Koll} \& {Abbot}(2015)}]{koll2015phasecurve}
{Koll}, D.~D.~B., \& {Abbot}, D.~S. 2015, \apj, 802, 21,
  \dodoi{10.1088/0004-637X/802/1/21}

\bibitem[{{Koll} \& {Abbot}(2016)}]{koll2016heat}
---. 2016, \apj, 825, 99, \dodoi{10.3847/0004-637X/825/2/99}

\bibitem[{{Komacek} {et~al.}(2020){Komacek}, {Fauchez}, {Wolf}, \&
  {Abbot}}]{komacek2020cloud}
{Komacek}, T.~D., {Fauchez}, T.~J., {Wolf}, E.~T., \& {Abbot}, D.~S. 2020,
  \apjl, 888, L20, \dodoi{10.3847/2041-8213/ab6200}

\bibitem[{{Kopparapu} {et~al.}(2016){Kopparapu}, {Wolf}, {Haqq-Misra}, {Yang},
  {Kasting}, {Meadows}, {Terrien}, \& {Mahadevan}}]{kopparapu2016innerhz}
{Kopparapu}, R.~k., {Wolf}, E.~T., {Haqq-Misra}, J., {et~al.} 2016, \apj, 819,
  84, \dodoi{10.3847/0004-637X/819/1/84}

\bibitem[{{Laskar} {et~al.}(2002){Laskar}, {Levrard}, \&
  {Mustard}}]{laskar2002mars}
{Laskar}, J., {Levrard}, B., \& {Mustard}, J.~F. 2002, \nat, 419, 375,
  \dodoi{10.1038/nature01066}

\bibitem[{{Leconte} {et~al.}(2013){Leconte}, {Forget}, {Charnay}, {Wordsworth},
  {Selsis}, {Millour}, \& {Spiga}}]{leconte2013bistable}
{Leconte}, J., {Forget}, F., {Charnay}, B., {et~al.} 2013, \aap, 554, A69,
  \dodoi{10.1051/0004-6361/201321042}

\bibitem[{{Loftus} \& {Wordsworth}(2021)}]{loftus2021rain}
{Loftus}, K., \& {Wordsworth}, R. 2021, arXiv e-prints, arXiv:2102.09570.
\newblock \doarXiv{2102.09570}

\bibitem[{{Luger} \& {Barnes}(2015)}]{luger2015waterloss}
{Luger}, R., \& {Barnes}, R. 2015, Astrobiology, 15, 119,
  \dodoi{10.1089/ast.2014.1231}

\bibitem[{{MacKenzie} {et~al.}(2019){MacKenzie}, {Barnes}, {Hofgartner},
  {Birch}, {Hedman}, {Lucas}, {Rodriguez}, {Turtle}, \&
  {Sotin}}]{mackenzie2019titanlake}
{MacKenzie}, S.~M., {Barnes}, J.~W., {Hofgartner}, J.~D., {et~al.} 2019, Nature
  Astronomy, 3, 506, \dodoi{10.1038/s41550-018-0687-6}

\bibitem[{{Merlis} \& {Schneider}(2010)}]{merlis2010tidally}
{Merlis}, T.~M., \& {Schneider}, T. 2010, Journal of Advances in Modeling Earth
  Systems, 2, 13, \dodoi{10.3894/JAMES.2010.2.13}

\bibitem[{{Mitchell} \& {Lora}(2016)}]{mitchell2016titan}
{Mitchell}, J.~L., \& {Lora}, J.~M. 2016, Annual Review of Earth and Planetary
  Sciences, 44, 353, \dodoi{10.1146/annurev-earth-060115-012428}

\bibitem[{{Montmessin} {et~al.}(2017){Montmessin}, {Smith}, {Langevin},
  {Mellon}, \& {Fedorova}}]{montmessin2017marswater}
{Montmessin}, F., {Smith}, M.~D., {Langevin}, Y., {Mellon}, M.~T., \&
  {Fedorova}, A. 2017, {The Water Cycle. The atmosphere and climate of Mars, R.
  M. Haberle, R. T. Clancy, F. Forget, M. D. Smith, and R. W. Zurek }
  (Cambridge University Press), 295--337, \dodoi{10.1017/9781139060172.011}

\bibitem[{{Noda} {et~al.}(2017){Noda}, {Ishiwatari}, {Nakajima}, {Takahashi},
  {Takehiro}, {Onishi}, {Hashimoto}, {Kuramoto}, \&
  {Hayashi}}]{noda2017rotation}
{Noda}, S., {Ishiwatari}, M., {Nakajima}, K., {et~al.} 2017, \icarus, 282, 1,
  \dodoi{10.1016/j.icarus.2016.09.004}

\bibitem[{{O'Gorman} \& {Schneider}(2008)}]{ogorman2008water}
{O'Gorman}, P.~A., \& {Schneider}, T. 2008, Journal of Climate, 21, 3815,
  \dodoi{10.1175/2007JCLI2065.1}

\bibitem[{{Pidhorodetska} {et~al.}(2020){Pidhorodetska}, {Fauchez},
  {Villanueva}, {Domagal-Goldman}, \&
  {Kopparapu}}]{Pidhorodetska2020trappist1e}
{Pidhorodetska}, D., {Fauchez}, T.~J., {Villanueva}, G.~L., {Domagal-Goldman},
  S.~D., \& {Kopparapu}, R.~K. 2020, \apjl, 898, L33,
  \dodoi{10.3847/2041-8213/aba4a1}

\bibitem[{{Pierrehumbert}(1995)}]{pierrehumbert_thermostats_1995}
{Pierrehumbert}, R.~T. 1995, Journal of Atmospheric Sciences, 52, 1784,
  \dodoi{10.1175/1520-0469(1995)052<1784:TRFATL>2.0.CO;2}

\bibitem[{{Pierrehumbert}(1998)}]{pierrehumbert1998subtropical}
---. 1998, \grl, 25, 151, \dodoi{10.1029/97GL03563}

\bibitem[{{Pierrehumbert}(2010)}]{pierrehumbert2010principle}
---. 2010, {Principles of Planetary Climate} ({Cambridge University Press})

\bibitem[{Pierrehumbert {et~al.}(2007)Pierrehumbert, Brogniez, \&
  Roca}]{pierrehumbert2007relative}
Pierrehumbert, R.~T., Brogniez, H., \& Roca, R. 2007, On the relative humidity
  of the atmosphere. The Global Circulation of the Atmosphere, T. Schneider and
  A. Sobel, Eds,  Princeton University Press

\bibitem[{{Pierrehumbert} \& {Ding}(2016)}]{pierrehumbert2016nondilute}
{Pierrehumbert}, R.~T., \& {Ding}, F. 2016, Proceedings of the Royal Society of
  London Series A, 472, 20160107, \dodoi{10.1098/rspa.2016.0107}

\bibitem[{{Pierrehumbert} \& {Hammond}(2019)}]{pierrehumbert2019tidelocked}
{Pierrehumbert}, R.~T., \& {Hammond}, M. 2019, Annual Review of Fluid
  Mechanics, 51, 275, \dodoi{10.1146/annurev-fluid-010518-040516}

\bibitem[{{Ramirez} \& {Kaltenegger}(2014)}]{ramirez2014premain}
{Ramirez}, R.~M., \& {Kaltenegger}, L. 2014, \apjl, 797, L25,
  \dodoi{10.1088/2041-8205/797/2/L25}

\bibitem[{{Suissa} {et~al.}(2020){Suissa}, {Mandell}, {Wolf}, {Villanueva},
  {Fauchez}, \& {Kopparapu}}]{suissa2020dim}
{Suissa}, G., {Mandell}, A.~M., {Wolf}, E.~T., {et~al.} 2020, \apj, 891, 58,
  \dodoi{10.3847/1538-4357/ab72f9}

\bibitem[{{Tian} \& {Ida}(2015)}]{tian2015waterloss}
{Tian}, F., \& {Ida}, S. 2015, Nature Geoscience, 8, 177,
  \dodoi{10.1038/ngeo2372}

\bibitem[{{Wordsworth}(2015)}]{wordsworth2015heat}
{Wordsworth}, R. 2015, \apj, 806, 180, \dodoi{10.1088/0004-637X/806/2/180}

\bibitem[{{Wordsworth} {et~al.}(2013){Wordsworth}, {Forget}, {Millour}, {Head},
  {Madeleine}, \& {Charnay}}]{wordsworth2013mars}
{Wordsworth}, R., {Forget}, F., {Millour}, E., {et~al.} 2013, \icarus, 222, 1,
  \dodoi{10.1016/j.icarus.2012.09.036}

\bibitem[{{Wordsworth} {et~al.}(2021){Wordsworth}, {Knoll}, {Hurowitz}, {Baum},
  {Ehlmann}, {Head}, \& {Steakley}}]{wordsworth2021episodic}
{Wordsworth}, R., {Knoll}, A.~H., {Hurowitz}, J., {et~al.} 2021, Nature
  Geoscience, 14, 127, \dodoi{10.1038/s41561-021-00701-8}

\bibitem[{{Wordsworth} {et~al.}(2011){Wordsworth}, {Forget}, {Selsis},
  {Millour}, {Charnay}, \& {Madeleine}}]{wordsworth2011gj581d}
{Wordsworth}, R.~D., {Forget}, F., {Selsis}, F., {et~al.} 2011, \apjl, 733,
  L48, \dodoi{10.1088/2041-8205/733/2/L48}

\bibitem[{{Yang} \& {Abbot}(2014)}]{yang2014twocolumn}
{Yang}, J., \& {Abbot}, D.~S. 2014, \apj, 784, 155,
  \dodoi{10.1088/0004-637X/784/2/155}

\bibitem[{{Yang} {et~al.}(2013){Yang}, {Cowan}, \& {Abbot}}]{yang2013innerhz}
{Yang}, J., {Cowan}, N.~B., \& {Abbot}, D.~S. 2013, \apj, 771, L45,
  \dodoi{10.1088/2041-8205/771/2/L45}

\bibitem[{{Yang} {et~al.}(2019){Yang}, {Leconte}, {Wolf}, {Merlis}, {Koll},
  {Forget}, \& {Abbot}}]{yang2019a3dcomparison}
{Yang}, J., {Leconte}, J., {Wolf}, E.~T., {et~al.} 2019, \apj, 875, 46,
  \dodoi{10.3847/1538-4357/ab09f1}

\bibitem[{{Zhu} \& {Newell}(1994)}]{zhu1994atmriver}
{Zhu}, Y., \& {Newell}, R.~E. 1994, \grl, 21, 1999, \dodoi{10.1029/94GL01710}

\bibitem[{{Zhu} \& {Newell}(1998)}]{zhu1998atmriver}
---. 1998, Monthly Weather Review, 126, 725

\end{thebibliography}
\bibliographystyle{aasjournal}


\listofchanges

\end{document}